\documentstyle[twoside,fleqn,espcrc2,epsfig]{article}
\newcommand{\be}{\begin{equation}}
\newcommand{\ee}{\end{equation}}
\newcommand{\ba}{\begin{eqnarray}}
\newcommand{\ea}{\end{eqnarray}}

\makeatletter
\newbox\slashbox \setbox\slashbox=\hbox{\large$/$}
\def\pslash#1{\setbox\@tempboxa=\hbox{$#1$}
  \@tempdima=0.5\wd\slashbox \advance\@tempdima 0.5\wd\@tempboxa
  \copy\slashbox \kern-\@tempdima \box\@tempboxa}
\def\FMSlash{\protect\pslash}
\title{Quantum chaos and chiral symmetry at the QCD and QED phase transition}
\author{Elmar Bittner, Harald Markum, and Rainer 
Pullirsch
\address{Institute for Nuclear Physics, TU-Wien,
Wiedner Hauptstra\ss e 8-10, A-1040 Vienna, Austria}}
\begin{document}
\begin{abstract} 
 We investigate the eigenvalue spectrum of the staggered Dirac matrix
  in SU(3) gauge theory and in full QCD as well as in quenched U(1)
  theory.  As a measure of the fluctuation
  properties of the eigenvalues, we consider the nearest-neighbor spacing
  distribution.
  We find that in all regions of their phase diagrams,
  compact lattice gauge theories have bulk spectral correlations given
  by random matrix theory, which is an indication for quantum chaos.
  In the confinement phase, the low-lying Dirac spectrum of these quantum field
  theories is well described by random matrix theory, exhibiting universal
  behavior. Related results for gauge theories with minimal coupling are now
  discussed also in the chirally symmetric phase.
\end{abstract}
\date{\today}
\maketitle

\section{Quantum chaos}

\begin{figure*}[t]
  \centerline{\psfig{figure=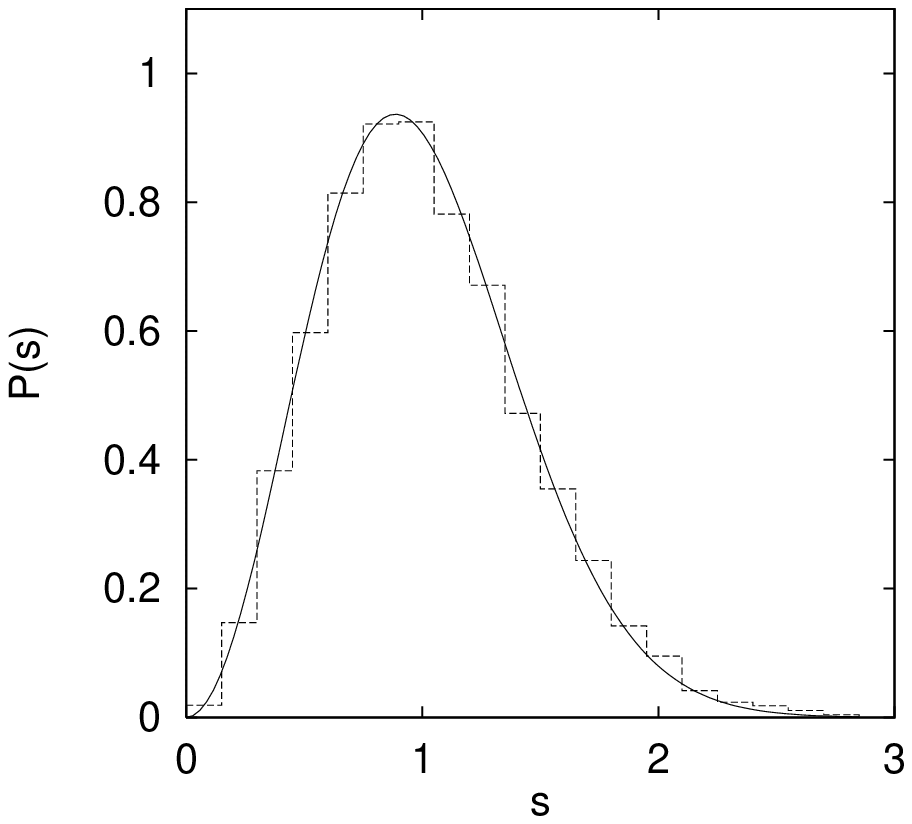,width=5cm}\hspace*{10mm}
    \psfig{figure=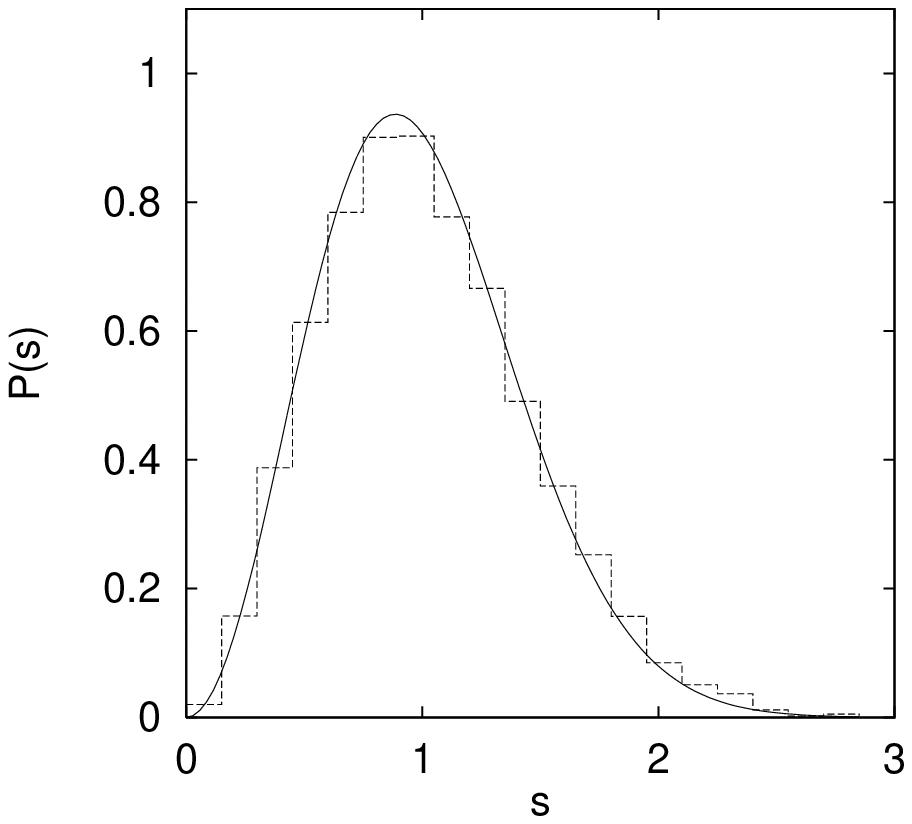,width=5cm}}
\vspace*{-10mm}
\caption{Nearest-neighbor spacing distribution $P(s)$ for full QCD
  on a $6^3 \times 4$ lattice in the confinement phase (left) and
  in the deconfinement phase (right) compared with the
  random matrix result (solid lines). There are no changes in $P(s)$
  across the deconfinement phase transition.}
\label{fintemp}
\vspace*{5mm}
  \centerline{\psfig{figure=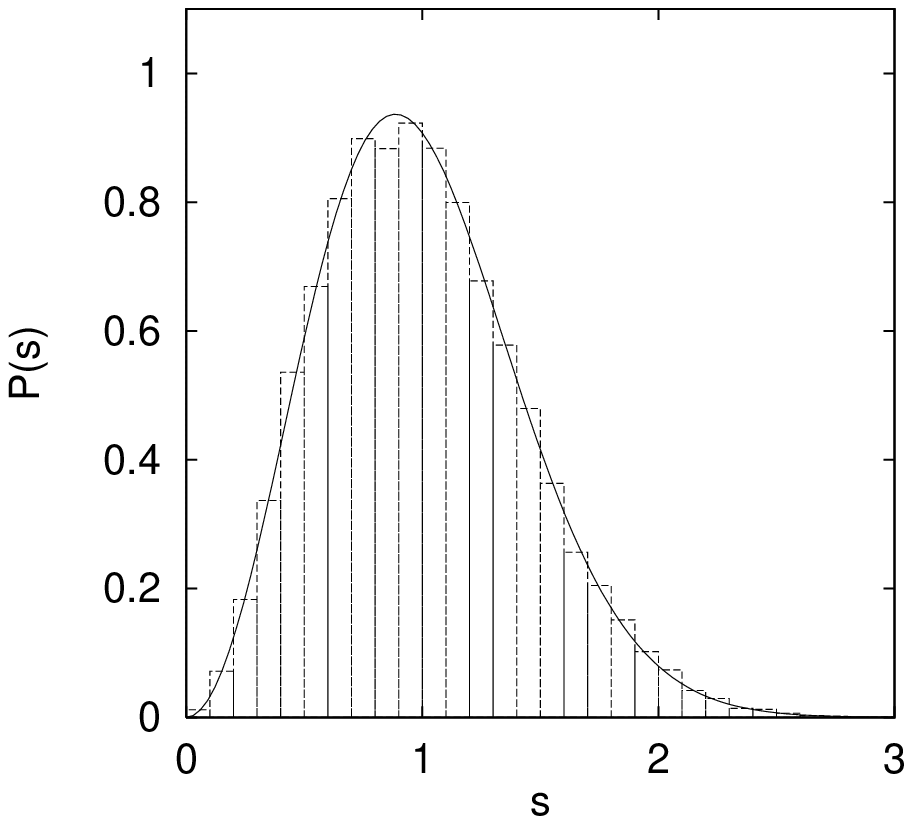,width=5cm}\hspace*{10mm}
    \psfig{figure=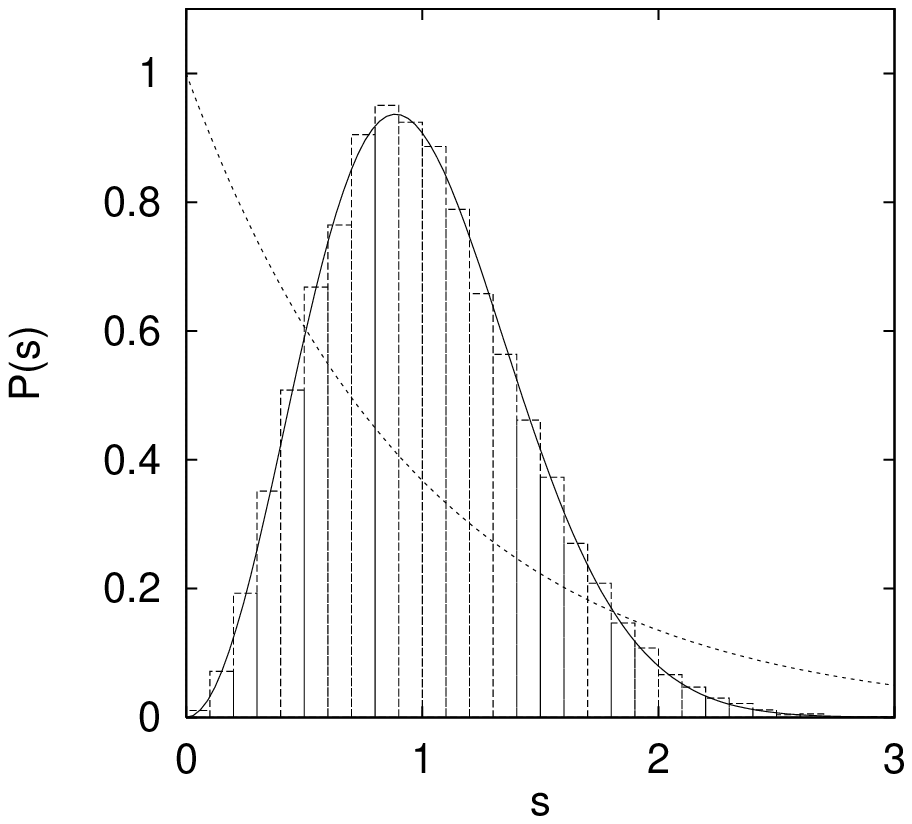,width=5cm}}
\vspace*{-10mm}
  \caption{Nearest-neighbor spacing distribution $P(s)$ for U(1) gauge
    theory on an $8^3\times 6$ lattice in the confined phase (left)
    and in the Coulomb phase (right). The theoretical curves are the chUE
    result, Eq.~(\ref{wigner}), and the Poisson distribution, 
    $P_{\rm P}(s)=e^{-s}$.}
  \label{f02}
\vspace{-0.5mm}
\end{figure*}

The eigenvalues of the Dirac operator are of great
interest for the universality of important features of QCD and QED.  On
the one hand, the accumulation of small eigenvalues is, via the
Banks-Casher formula~\cite{Bank80}, related to the spontaneous
breaking of chiral symmetry. On the other hand, the fluctuation
properties of the eigenvalues in the bulk of the spectrum
can be described by random matrix theory (RMT), see Ref.~\cite{Hala95}.
For example, the nearest-neighbor spacing
distribution $P(s)$, i.e., the distribution of spacings $s$ between
adjacent eigenvalues on the unfolded scale, agrees with the Wigner
surmise of RMT.  According to the Bohigas-Giannoni-Schmit
conjecture~\cite{Bohi84}, quantum systems whose classical counterparts are
chaotic~\cite{Biro99} have a nearest-neighbor spacing distribution given by RMT
whereas systems whose classical counterparts are integrable obey a
Poisson distribution, $P_{\rm P}(s)=e^{-s}$.  Therefore, the specific
form of $P(s)$ is often taken as a criterion for the presence or
absence of ``quantum chaos''.

In RMT, one has to distinguish several universality classes which are
determined by the symmetries of the system.  For the case of the QCD
Dirac operator, this classification was done in
Ref.~\cite{Verb94}.  Depending on the number of colors and the
representation of the quarks, the Dirac operator is described by one
of the three chiral ensembles of RMT.  As far as the fluctuation
properties in the bulk of the spectrum are concerned, the predictions
of the chiral ensembles are identical to those of the ordinary
ensembles~\cite{Fox64}.  In Ref.~\cite{Hala95}, the Dirac matrix
was studied for color-SU(2) using both staggered and Wilson fermions which
correspond to the chiral symplectic (chSE) and orthogonal (chOE) ensemble,
respectively.  Here~\cite{Pull98}, we study SU(3) with staggered
fermions which corresponds to the chiral unitary ensemble (chUE). The
RMT result for the nearest-neighbor spacing distribution can be
expressed in terms of so-called prolate spheroidal functions, see
Ref.~\cite{Meht91}.  A very good approximation to $P(s)$ is
provided by the Wigner surmise for the unitary ensemble,
\begin{equation} \label{wigner}
  P_{\rm W}(s)=\frac{32}{\pi^2}s^2e^{-4s^2/\pi} \:.
\end{equation}

We generated gauge field configurations using the standard Wilson
plaquette action for SU(3) with and without dynamical fermions in the
Kogut-Susskind prescription. We have worked on a $6^3\times 4$ lattice
with various values of the inverse gauge coupling $\beta=6/g^2$ both
in the confinement and deconfinement phase.  We typically produced 10
independent equilibrium configurations for each $\beta$.  

The Dirac operator, $\FMSlash{D}=\FMSlash{\partial}+ig\FMSlash{A}$, is
anti-Hermitian so that the eigenvalues $\lambda_n$ of $i\FMSlash{D}$
are real.  Because of $\{\FMSlash{D},\gamma_5\}=0$ the nonzero
$\lambda_n$ occur in pairs of opposite sign.  All spectra were checked
against the analytical sum rules $\sum_{n} \lambda_n = 0$ and
$\sum_{\lambda_n>0} \lambda_n^2 = 3V$, where V is the lattice volume.
To construct the nearest-neighbor spacing distribution from the
eigenvalues, one first has to ``unfold'' the spectra~\cite{Meht91}.

Figure~\ref{fintemp} compares $P(s)$ of full QCD with $N_f = 3$
flavors and quark mass $ma=0.05$ to the RMT result.  In the confinement
$(\beta=5.2)$ as well as in the deconfinement $(\beta=5.4)$ phase we
observe agreement with RMT up
to very high $\beta$ (not shown).  The observation that $P(s)$ is not
influenced by the presence of dynamical quarks is
expected from the results of Ref.~\cite{Fox64}, which
apply to the case of massless quarks. Our
results, and those of Ref.~\cite{Hala95}, indicate that massive
dynamical quarks do not affect $P(s)$ either.

We have also investigated the staggered Dirac spectrum of 4d U(1)
gauge theory which corresponds to the chUE of RMT. At $\beta_c \approx 1.01$
U(1) gauge theory undergoes a phase transition between a confinement
phase with mass gap and monopole excitations for $\beta < \beta_c$ and
the Coulomb phase which exhibits a massless photon for $\beta >
\beta_c$~\cite{BePa84}. As for SU(2) and SU(3) gauge groups, we
expect the confined phase to be described by RMT, whereas free
fermions yield the Poisson distribution.
The question arose whether the Coulomb phase would be
described by RMT or by the Poisson distribution~\cite{BeMaPu99}.  The
nearest-neighbor spacing distributions for an $8^3\times 6$ lattice at
$\beta=0.9$ (confined phase) and at $\beta=1.1$ (Coulomb phase),
averaged over 20 independent configurations, are depicted in
Fig.~\ref{f02}. Both are well described by the chUE of RMT.

\begin{figure}[h]
  \centerline{\psfig{figure=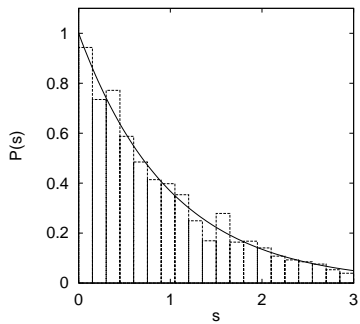,width=4.75cm}}
\vspace{-10mm}
  \caption{Nearest-neighbor spacing distribution $P(s)$ for the free
            Dirac operator on a $53\times 47\times 43\times 41$ lattice
            compared with a Poisson distribution.}
  \label{free}
\end{figure}

No signs for a transition to Poisson regularity are found. The
deconfinement phase transition does not seem to coincide with a
transition in the spacing distribution. For very large values of
$\beta$ far into the deconfinement region, the eigenvalues
start to approach the degenerate eigenvalues of the free theory, given
by $\lambda^2=\sum_{\mu=1}^4 \sin^2(2\pi n_\mu/L_\mu)/a^2$, where $a$
is the lattice constant, $L_{\mu}$ is the number of lattice sites in
the $\mu$-direction, and $n_\mu=0,\ldots,L_\mu-1$.  In this case, the
spacing distribution is neither Wigner nor Poisson.
It is possible to lift the degeneracies of the free
eigenvalues using an asymmetric lattice where $L_x$, $L_y$, etc. are
relative primes and, for large lattices, the distribution
is then Poisson, $P_{\rm P}(s)=e^{-s}$, see Fig.~\ref{free}.
\vspace{-3mm}

\vspace{-8mm}
\section{Chiral symmetry}

\begin{figure*}[p]
    \centerline{\psfig{figure=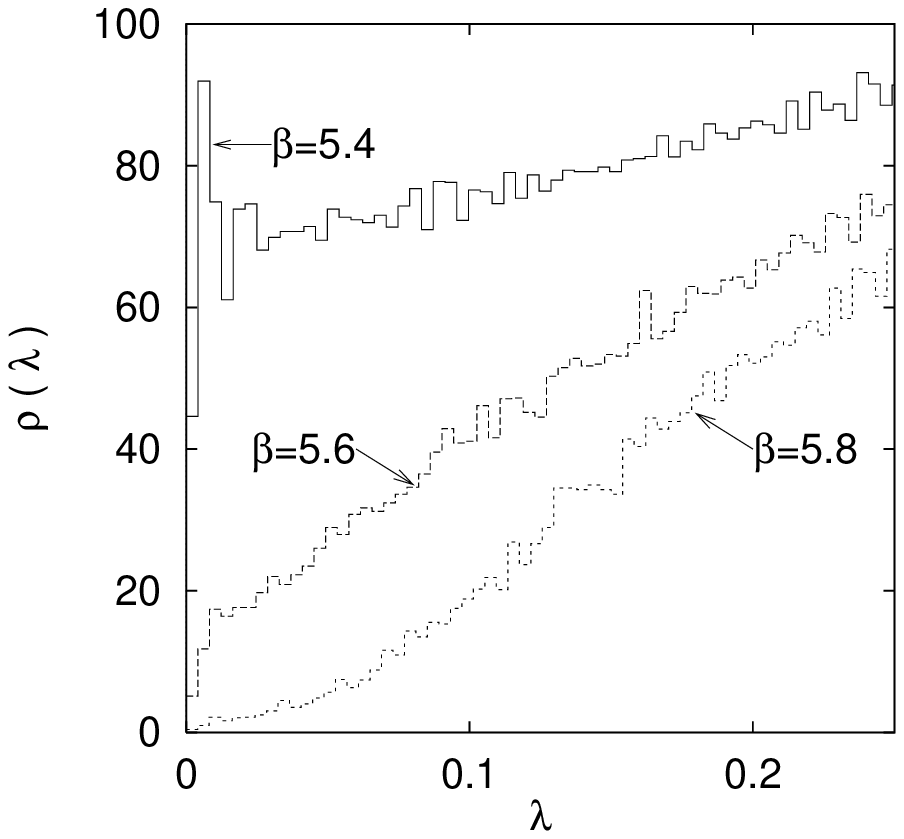,width=5cm}\hspace*{10mm}
                \psfig{figure=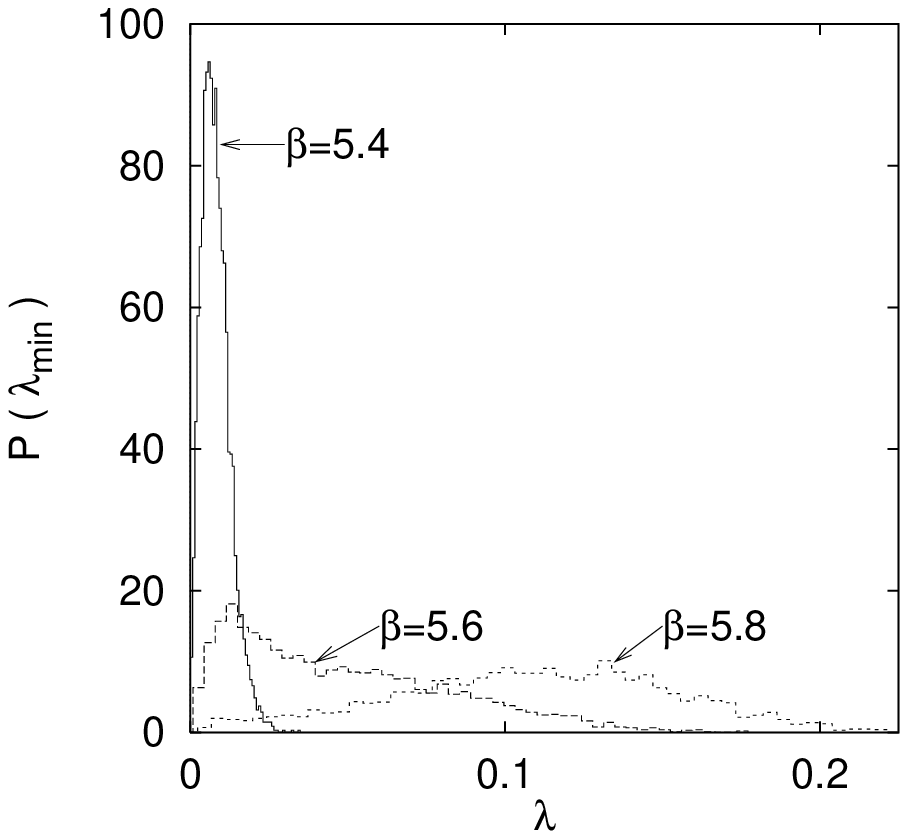,width=5cm}}
\vspace*{-10mm}
  \caption{Density $\rho (\lambda)$ of small eigenvalues (left) and distribution
    $P(\lambda_{\min})$ (right) for SU(3) gauge theory on a $4^4$ lattice across
    the phase transition.}
  \label{fig4}
\vspace*{5mm}
  \centerline{\psfig{figure=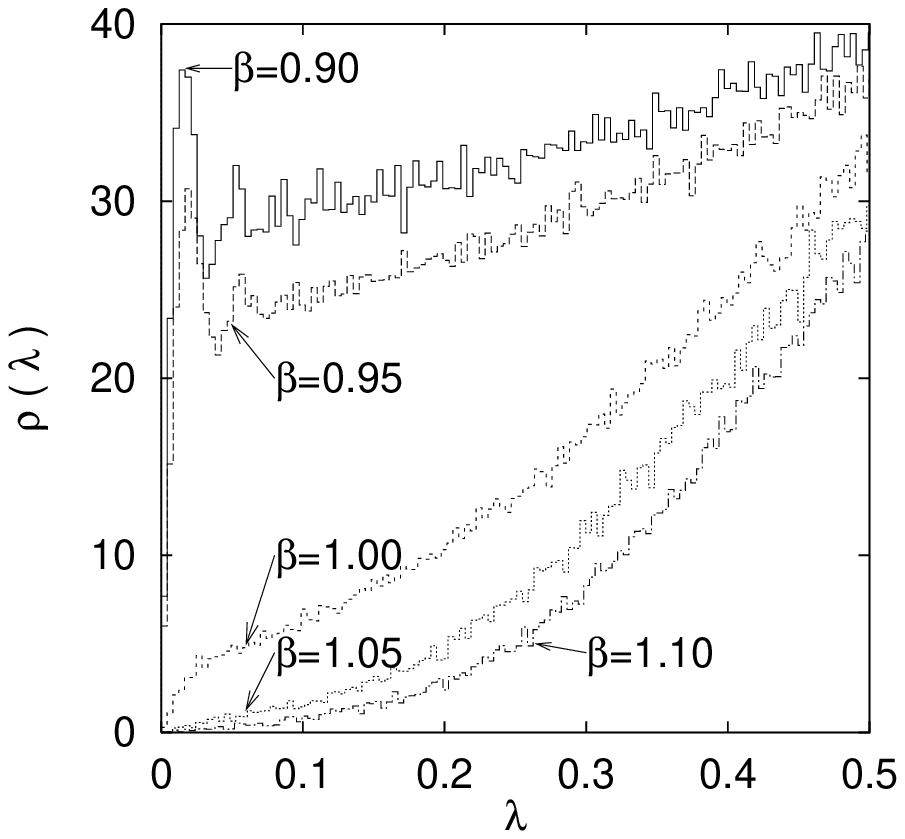,width=5cm}\hspace*{10mm}
  \psfig{figure=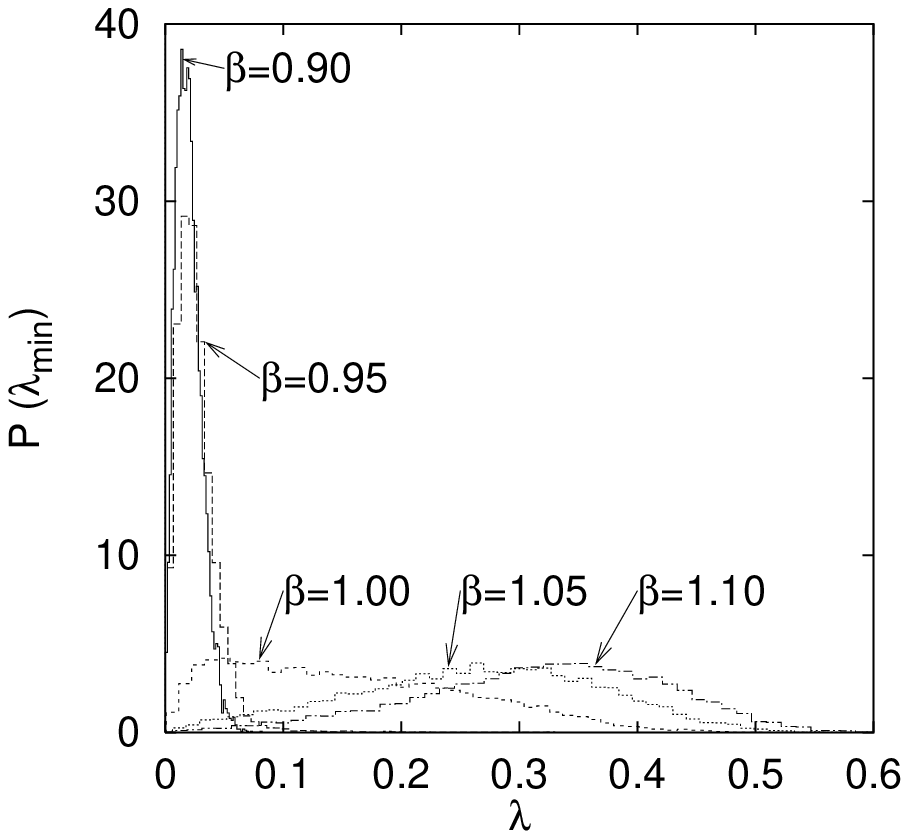,width=5cm}}
\vspace*{-10mm}
  \caption{Density $\rho (\lambda)$ of small eigenvalues (left) and distribution
    $P(\lambda_{\min})$ (right) for U(1) gauge theory on a $4^4$ lattice across
    the phase transition.}
  \label{fig5}
\vspace*{5mm}
  \centerline{\psfig{figure=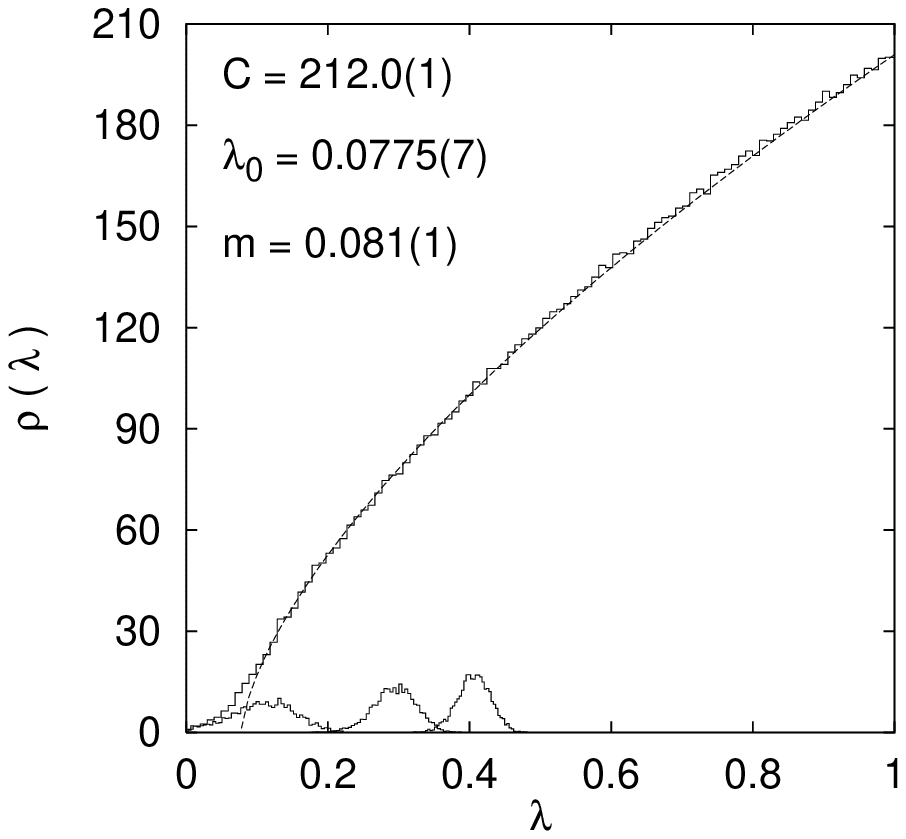,width=5cm}\hspace*{10mm}
  \psfig{figure=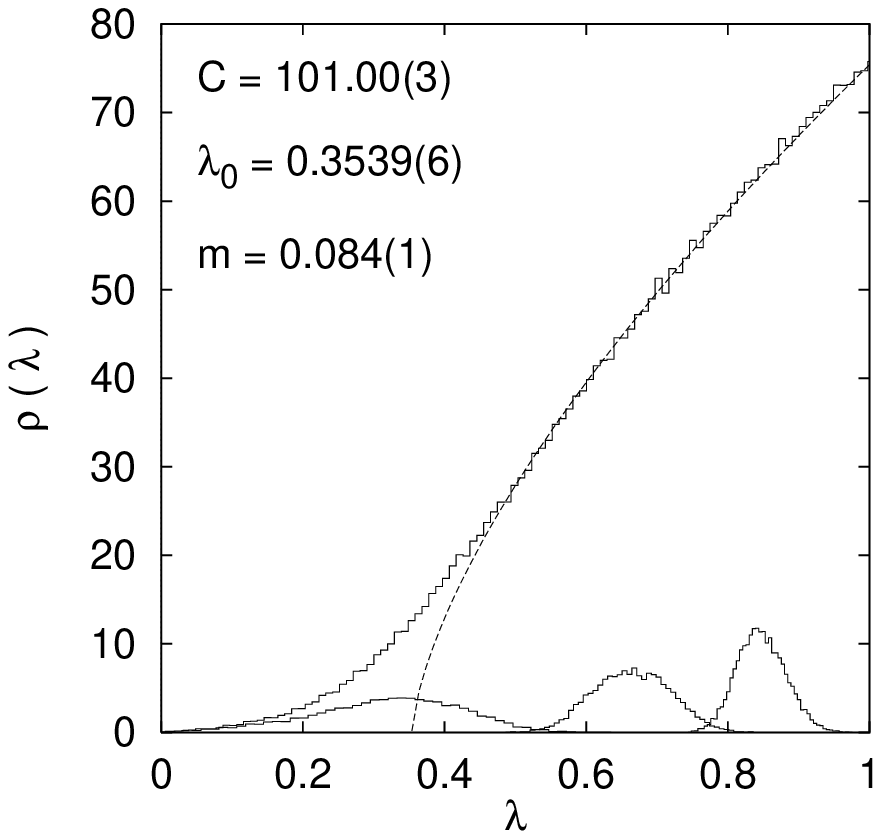,width=5cm}}
\vspace*{-10mm}
  \caption{Fit of the spectral density to $\rho (\lambda) =
    C (\lambda -\lambda_0)^{2m+1/2}$ in SU(3) at $\beta=5.8$ (left)
    and in U(1) at $\beta=1.10$ (right). The contribution
    of the smallest eigenvalue, the $11^{{\mbox{th}}}$ eigenvalue and the
    $21^{{\mbox{st}}}$ eigenvalue is inserted.}
  \label{fig6}
\end{figure*}

We have continued our investigations with a study of the
distribution of the small eigenvalues in the whole phase diagram. The
Banks-Casher formula \cite{Bank80} relates the Dirac eigenvalue density
$\rho(\lambda)$ at $\lambda=0$ to the chiral condensate,
$ \Sigma \equiv |\langle \bar{\psi} \psi \rangle| =
 \lim_{\varepsilon\to 0}\lim_{V\to\infty} \pi\rho (\varepsilon)/V$.
The microscopic spectral density,
$ \rho_s (z) = \lim_{V\to\infty}
 \rho \left( {z/V\Sigma } \right)/V\Sigma , $
should be given by the appropriate prediction of RMT~\cite{ShVe92}, which
also generates the Leutwyler-Smilga sum rules~\cite{LeSm92}.

We present results in Fig.~\ref{fig4} for SU(3) theory and the
staggered Dirac operator on a $4^4$ lattice from 5000 configurations
for $\beta=5.4$ and 3000 configurations for $\beta=5.6$ and $\beta=5.8$
around the critical temperature $\beta_c \approx 5.7$. In the confinement
phase, both the microscopic spectral density $\rho_s(z)$ and the 
distribution $P(\lambda_{\rm min})$ of the smallest eigenvalue
agree with the RMT predictions of the chUE for topological charge $\nu = 0$
\cite{Goec99}.

Our analog presentation for U(1) theory is from 10000 configurations on a $4^4$
lattice around the critical coupling. In the left plot of
Fig.~\ref{fig5} a comparison with RMT for the microscopic
spectral density $\rho_s (z)$ yields again quite satisfactory agreement
in the confinement. The analytical RMT result for the
(quenched) chUE and $\nu=0$ is given by \cite{ShVe92}
$ \rho_s(z) = z\, [ J_0^2(z) + J_1^2(z) ]/2$, where $J$ denotes the
Bessel function.
The chiral condensate $\Sigma$ can be obtained by extrapolating the
histogram for $\rho(\lambda)$ to $\lambda=0$ and using the
Banks-Casher relation~\cite{Berg00}.
The right plot in Fig.~\ref{fig5}
exhibits the distribution $P(\lambda_{\min})$ of the
smallest eigenvalue, being in the chirally broken phase in accordance
with the prediction
of the (quenched) chUE of RMT for topological charge $\nu=0$,
$P(\lambda_{\min}) = (V\Sigma)^2 (\lambda_{\min}/2)\,\exp( -
(V\Sigma\lambda_{\min})^2/4)$.

The quasi-zero modes which are responsible for the chiral condensate
$\Sigma \neq 0$ build up when we cross from the deconfinement into the
confined phase.
Figures~\ref{fig4} and~\ref{fig5} demonstrate that both $\rho(\lambda)$
and $P(\lambda_{min})$ plotted with varying $\beta$ on identical scales,
respectively, can serve as an indicator for the phase transition.

In Fig.~\ref{fig6} we turn to a discussion of the spectrum in the 
quark-gluon plasma and Coulomb phase. From RMT a functional form of
$\rho (\lambda) = C (\lambda -\lambda_0)^{2m+1/2}$ is expected at
the onset of the eigenvalue density~\cite{Bowi91}. A fit to the data
in the regime up to $\lambda = 1$ yields
$m=0.081(1)$ for the non-Abelian and $m=0.084(1)$ for the Abelian theory,
in agreement with recent studies~\cite{Lang99}. This suggests that
both theories correspond to universality class $m=0$. For this class
a microscopic level density involving the Airy function can be deduced
from RMT~\cite{Forr93}. A rescaling of our data from the $4^4$ lattice
to this functional form is not satisfactory for both theories~\cite{Lang99}.
The reason might be (i) that the lowest eigenvalue is still influenced
by quasi-zero modes from the confinement and (ii) that already the first
eigenvalues lie above an analogue of the Thouless energy. With increasing lattice
size these effects should decrease. Further, we checked that the high-end of the
spectrum behaves similar to the low-end and thus deviates from the
results of the ordinary UE of RMT~\cite{Forr93}. Again the question arises
whether larger lattices possess spectral edges closer to the microscopic semi-circle
universality.

\vspace{-3mm}
\section{Conclusions}

The aim of this contribution was to work out two different types of
universalities inherent in quantum field theories with a covariant
derivative and their
interpretation in terms of RMT. The first type concerns the bulk of
the spectrum of the Dirac operator.  The nearest-neighbor spacing
distribution $P(s)$ agrees with the RMT result in both the confinement
and the deconfinement phase of pure gauge theory and of full QCD,
except for extremely large values of $\beta$ where the eigenvalues are
known analytically.
The nearest-neighbor spacing distribution of 4d U(1) quenched lattice
gauge theory is described by the chUE of RMT in both the confinement
and the Coulomb phase.
Even in the deconfinement phase,
gauge fields retain a considerable degree of randomness, which
apparently gives rise to quantum chaos in these theories.

The second type of universality concerns the low-lying spectra of the
Dirac operators of both QCD and QED. In all cases considered, one
finds that in the phase in which  chiral symmetry is spontaneously
broken the distribution
$P(\lambda_{\min})$ and the microscopic spectral density $\rho_s(z)$
are described by chiral RMT. When chiral symmetry is restored one has
to rely on ordinary RMT with the related space-time symmetries, but
one finds for our lattice size only universal behavior of the
macroscopic density $\rho(\lambda)$.

{\it Acknowledgments:}
This study was supported in part by FWF project P11456-PHY.
We thank B.A. Berg, M.-P. Lombardo, and T. Wettig for 
collaborations~\cite{9900}.


\end{document}